\documentclass[aps,showpacs,twocolumn,english,prl]{revtex4}
\usepackage[T1]{fontenc}
\usepackage[latin9]{inputenc}
\usepackage{amsmath}
\usepackage{amssymb}
\usepackage{graphicx}
\usepackage{esint}
\usepackage{color}

\makeatletter
\@ifundefined{textcolor}{}
{%
 \definecolor{BLACK}{gray}{0}
 \definecolor{WHITE}{gray}{1}
 \definecolor{RED}{rgb}{1,0,0}
 \definecolor{GREEN}{rgb}{0,1,0}
 \definecolor{BLUE}{rgb}{0,0,1}
 \definecolor{CYAN}{cmyk}{1,0,0,0}
 \definecolor{MAGENTA}{cmyk}{0,1,0,0}
 \definecolor{YELLOW}{cmyk}{0,0,1,0}
}

\makeatother

\usepackage{babel}
\begin{document}
\global\long\def\ket#1{\left|#1\right\rangle }

\global\long\def\bra#1{\left\langle #1\right|}

\global\long\def\braket#1#2{\left\langle #1\left|#2\right.\right\rangle }

\global\long\def\ketbra#1#2{\left|#1\right\rangle \left\langle #2\right|}

\global\long\def\braOket#1#2#3{\left\langle #1\left|#2\right|#3\right\rangle }

\global\long\def\mc#1{\mathcal{#1}}

\global\long\def\nrm#1{\left\Vert #1\right\Vert }

\title{Purity and entropy evolution speed limits for open quantum systems}

\author{Raam Uzdin$^{1}$}

\author{Eric Lutz$^{2}$}

\author{Ronnie Kosloff$^{1}$ }

\address{$^{1}$Fritz Haber Research Center for Molecular Dynamics, Hebrew
University of Jerusalem, Jerusalem 91904, Israel}

\address{$^{2}$Department of Physics, Friedrich-Alexander-Universit\"at Erlangen-N\"urnberg, 91058 Erlangen, Germany}


\begin{abstract}
We derive generic upper bounds on the rate of purity change and entropy increase for open quantum systems. These bounds depend solely on the generators of the nonunitary dynamics and are independent of  the particular states of the systems. They are thus  perfectly suited to investigate dephasing and thermalization processes of arbitrary systems. We apply these results to single and multiple dephasing channels, to a problem of  quantum  control in the presence of noise, and to cooling.
\end{abstract}

\maketitle

Determining the maximal rate of evolution of an open system is of crucial importance in quantum physics. Any quantum system unavoidably couples to external degrees of freedom (the environment) that lead to  loss of phase coherence and/or  to thermalization \cite{bre07}. In most applications, the main challenge is to minimize the effect of the environment. In quantum computing \cite{nie00} and coherent control \cite{sha03}, for example, it is vital to achieve low dephasing rates in order to protect the system against decoherence. On the other hand, there are  many instances where it is of advantage to maximize the influence of the surroundings. A case in point is the cooling of a quantum system and the preparation of pure states, where high cooling rates are sought after \cite{wan13}. Additionally, in quantum thermodynamics, a  bound on the rate of thermalization of a heat engine with its  reservoirs will limit its cycle time  and therefore put a restriction on maximal power output \cite{kos14}. 

Bounds on the rate of quantum evolution  are useful to assess if a process can be completed in a given time, without having to explicitly solve the (usually complicated) equations of motion \cite{fle73,bha83,ana90,vai91,uff93,bro03}. Quantum speed limits, defined as  the time derivative of a geometric quantity, for instance an angle between two states, are often introduced to characterize the maximal rate of evolution of a quantum  system.  Two prominent examples of quantum speed limits for closed systems are the Mandelstam-Tamm bound, $|d_t \theta| \leq \Delta E_\psi/\hbar$ \cite{man45}, and the Margolus-Levitin bound, $|d_t \theta| \leq \langle E\rangle_\psi/\hbar$ \cite{mar98}. Here $\Delta E_\psi$ is the energy width of the initial pure state $\psi(0)$, $\langle E\rangle _\psi$ its mean energy above the ground state, and $\theta = \arccos |\langle \psi(0)|\psi(t) \rangle|$  the geometric angle between initial and final states (we will set $\hbar=1$ in the sequel).  In the last years, the Mandelstam-Tamm and Margolus-Levitin bounds have been extended to mixed \cite{gio03} and driven \cite{pfe93,pfe95,def13} closed quantum systems as well as to open quantum systems \cite{tad13,cam13,def13a}.   These speed limits  depend explicitly on the state $\rho$ of the system and can be written in the general form $|d_t G(\rho)|\leq f(\rho,{\cal A})$, where $G(\rho)$ is a geometric quantity that characterizes the state of the system  and $f$ is a function of both the state  and  the generator ${\cal A}$  of the quantum evolution -- the generator of unitary  evolution is the Hamiltonian $H$ of a closed system, while the generator of nonunitary Markovian evolution of an open system may be given by a Lindblad generator $L$ \cite{bre07}. The dependence of the bound on the system state $\rho$ is especially useful when the evolution speed of different states are  to be compared, as for example  in quantum metrology when the optimal state for phase estimation is needed \cite{tad13,cam13}. 

Here we consider another class of quantum speed limits of the form $|d_t G(\rho)|\leq f({\cal A})$, where the function $f$ depends solely on the generator ${\cal A}$. Since these bounds do not refer to any particular state $\rho$, they present a clear separation between the geometric part $G(\rho)$ that describes  the state of the system and the kinetic part $f({\cal A})$ that is controlled by the generator. These bounds are particularly well suited to investigate the impact of an external environment on generic quantum systems whose states either  not known in detail, or are too complex to be determined exactly, e.g. those of an interacting many-body system. Bounds of this type have been studied for Hermitian \cite{LidarNormActionAndDistanceBound} and non-Hermitian \cite{uzdin100evoSpeed} Hamiltonians  that often appear  in systems with absorption as in optics.  Our aim in this paper is to provide state-independent speed limits for the purity ${\cal P}=\text{tr} \rho^2$ and the von Neumann entropy $S= - \text{tr} \rho \ln \rho$ of a generic   open quantum system, two quantities that are commonly used to quantify the action of the environment \cite{bre07}. In the following, we first illustrate our method in the usual Hilbert space of density matrices by deriving a bound for the purity speed in terms of the Hilbert-Schmidt norm of the Lindblad operators of the nonunitary dynamics. We then extend our approach to Liouville space (space of density "vectors" as described later on) and obtain  a tighter inequality in terms of the spectral norm of the corresponding Hamilton superoperator. We further considerably  improve the bound by introducing a purity deviation  that is obtained by subtracting the steady-state dynamics of the time-dependent system.  Remarkably, we show that this bound is  tight, for all states and at all times, for a single qubit dephasing channel. We finally derive a  speed limit for  the rate of entropy decrease, and apply our formalism to quantum control in the presence of noise and to cooling.

\textit{Purity bound in Hilbert space}. We begin by deriving an upper bound to the purity speed limit in the density matrix
formulation. It will be convenient to consider an integral  bound of the type,
\begin{equation}
\left|G(\rho_{f})-G(\rho_i)\right|\le\int_{t_{i}}^{t_{f}}f({\cal A}_t)dt,\label{eq: int gen form}
\end{equation}
which follows from the differential bound, $|d_t G(\rho)|\leq f({\cal A}_t)$, with  the inequality $|\int_{t_{i}}^{t_{f}}d_tG(\rho)|\le\int_{t_{i}}^{t_{f}}\left|d_tG(\rho)\right|$. This form is often more appealing as it relates the initial and final
states of interest without reference to the intermediate dynamics.
The right-hand side will typically have the form of an "action" integral \cite{uzdinNHresources}.

We consider a possibly driven $N$-level quantum system with Hamiltonian $H_\rho \in\mathbb{C}^{N\times N}$.  We describe the nonunitary time evolution of its density matrix $\rho$ by  a Markovian master equation of the Lindblad-type \cite{bre07},
\begin{equation}
d_t\rho=L_t(\rho)=i[H_\rho,\rho]+\sum_{k}A_{k}\rho A_{k}^{\dagger}-\{\frac{1}{2}A_{k}^{\dagger}A_{k},\rho\},\label{eq: Lindblad eq.}
\end{equation}
where the operators $A_{k}\in\mathbb{C}^{N\times N}$ describe the interaction
with the external environment ($L_t$ is the time dependent Lindblad generator of the nonunitary dynamics). Master equations of the form \eqref{eq: Lindblad eq.} are the tool of choice to investigate the dynamics of systems weakly coupled to a reservoir in quantum optics and solid state physics \cite{bre07}. According to  a theorem by Lidar, Shabani and Alicki, the purity for any   $\rho$  only decreases 
if and only if $[A_{k},A_{k}^{\dagger}]=0$ for all k \cite{LidarAlickiPurityDecrease}. This condition provides  useful means
to separate between purely dephasing processes that can only reduce
the purity and processes that have the capability reducing the entropy and cooling the system.

We  use the cyclic property of the trace to write  $d_t\ln \text{tr}(\rho^{2})={2\text{tr}(\rho L_t(\rho))}/{\text{tr}(\rho^{2})}$. Integrating over time and using the triangle inequality, we  have,
\begin{eqnarray}
\left|\ln\frac{\mc P(t_{f})}{\mc P(t_{i})}\right| & \le & \int\frac{2\left|\text{tr}(\rho L(\rho))\right|}{\text{tr}(\rho^{2})}dt.\label{eq: rho purity triang}
\end{eqnarray}
We next exploit the fact that $\mc P(t)= \text{tr}(\rho^{2})=\nrm{\rho}^{2}_2$, where  $\nrm{\cdot}_2$ denotes the Hilbert-Schmidt norm. An upper bound to Eq.~(\ref{eq: rho purity triang}) can be derived  with the help of elementary matrix algebra \cite{bha97}. Combining the Cauchy-Schwarz inequality, $|\text{tr}(\rho L_t(\rho))|\le\nrm{\rho}_2\nrm{L_t(\rho)}_2$,
the triangle inequality together with the submultiplicativity property of the norm and the master equation \eqref{eq: Lindblad eq.}, we find $\nrm{\rho}_2\nrm{L_t(\rho)}_2\le 2\sum_k \nrm{A_{k}}^{2}_2\nrm{\rho}^{2}_2$.
Inserting this expression into Eq.~(\ref{eq: rho purity triang}), we obtain  a "norm
action" integral inequality of  type (\ref{eq: int gen form}) for the logarithm of the purity:
\begin{equation}
\left|\ln\frac{\mc P(t_{f})}{\mc P(t_{i})}\right|\le4\int_{t_{i}}^{t_{f}}\sum_k\nrm{A_{k}}_2^{2}dt.\label{eq: H_rho purity bound}
\end{equation}
The quantity $-\ln\mc P$ is  known as   the Rényi entropy of order two or  the collision
entropy \cite{RenyiEntropyQM}. We note that the unitary evolution term,
 $i[\rho,H_\rho]$,  can  be  omitted, since the norm is unitarily invariant. Equation \eqref{eq: H_rho purity bound} provides an upper bound to the speed of variation of the Rényi entropy of order two of the system in terms of the Hilbert-Schmidt norm of the Lindblad operators.  Its practical usefulness stems from the fact that  the operators $A_k$ can be determined via quantum process tomography \cite{chu97}. This technique has been successfully demonstrated experimentally in NMR systems \cite{chi01}, solid state qubits \cite{how06}, vibrational states of atoms in optical lattices \cite{myr05}, quantum gate operations \cite{obr04,nam05,rie06} and quantum memory \cite{nee08}, as well as to  relaxing photon fields inside cavities \cite{bru08} and  superconducting quantum circuits \cite{wan08}. It is worth noticing that the purity has been measured directly in some cases, without having recourse to full quantum state tomography  \cite{bov05,du06,ada07}. However, it is notoriously difficult to measure nonlinear functions of the density operator \cite{eke02} and Eq.~\eqref{eq: H_rho purity bound} allows to estimate the rate of change of the Rényi entropy from the more accessible Lindblad operators.

\textit{Purity bounds in Liouville space}. Quantum dynamics is traditionally described in Hilbert space. However, it is sometimes convenient, in particular for open quantum systems, to introduce an extended space where  density operators are simple vectors and  time evolution is generated by superoperators. This space is usually referred to as Liouville space \cite{muk95}. We denote the "density vector" by $\ket r\in\mathbb{C}^{1\times N^{2}}$. It can be  obtained by reshaping the density matrix $\rho$ into a larger single vector with  index  $\alpha\in\{1,2,....N^{2}\}.$ The one-to-one mapping
of the two matrix indices into a single vector index $\{i,j\}\to\alpha$
is arbitrary, but has to be used consistently. The equation of motion
of the density vector in Liouville space follows from $d_t\rho_{\alpha}=\sum_{\beta} \rho_{\beta}\partial(d_t\rho_{\alpha})/\partial\rho_{\beta}$. Using this equation one can verify that the dynamics of the density vector $\ket r$ is governed by a Schr\"odinger-like equation in the new space,
\begin{equation}
i\partial_{t}\ket r=H_{r}\ket r, \label{eq: schrodinger eq}
\end{equation} 
where the  Hamiltonian superoperator $H_r \in\mathbb{C}^{N^{2}\times N^{2}}$ is given by,
\begin{equation}
H_{r,\alpha\beta}=i\frac{\partial(d_t\rho_{\alpha})}{\partial\rho_{\beta}}.\label{eq: Hr form}
\end{equation}
The vector $\ket r$ is not normalized to unity in general. Its  norm is  equal to the
purity, $\mc P=\text{tr}(\rho^{2})=\braket rr$,
where $\bra r=\ket r^{\dagger}$ as usual. It is important to note that not all vectors in Liouville space can be populated exclusively. This is due to the fact
that only positive $\rho$ with unit trace are legitimate density
matrices. The states that can be populated exclusively describe steady states, while the other ones correspond to
transient changes. For open systems, the Hamiltonian $H_{r}$ is in general non-Hermitian.
The skew Hermitian part $(H_{r}-H_{r}^{\dagger})/2$ is responsible
for purity changes and will thus play  a central role in the following. This
term originates uniquely from the Lindblad operators $A_{k}$ of the master equation \eqref{eq: Lindblad eq.}. 

We may now derive a purity bound in Liouville space by repeating the procedure previously used in Hilbert space. Starting from the Schr\"odinger-like equation \eqref{eq: schrodinger eq}, we first obtain the equality,
\begin{equation}
\frac{\partial_{t}\braket rr}{\braket rr}=-i\frac{\braOket r{H_{r}-H_{r}^{\dagger}}r}{\braket rr}.
\end{equation}
Integrating this  expression over time and using the triangle inequality, we get,
\begin{equation}
\left|\ln\frac{\mc P(t_f)}{\mc P(t_i)}\right|\le\int_{t_i}^{t_f}\frac{\left|\braOket r{H_{r}-H_{r}^{\dagger}}r\right|}{\braket rr}dt.\label{eq: r purity evo triang}
\end{equation}
The integrand may be further bounded by the spectral norm of the skew Hermitian part \cite{bha97},
\begin{equation}
\frac{\left|\braOket r{H_{r}-H_{r}^{\dagger}}r\right|}{\braket rr}\le\nrm{H_{r}-H_{r}^{\dagger}}_\text{sp}.\label{eq: spec norm speed bound}
\end{equation}
Combining Eqs.~(\ref{eq: r purity evo triang}) and (\ref{eq: spec norm speed bound}), we eventually obtain a norm action bound for the R\'eyni entropy   of the form (\ref{eq: int gen form}):

\begin{equation}
\left|\ln\frac{\mc P(t_f)}{\mc P(t_i)}\right|\le\int_{t_i}^{t_f}\nrm{H_{r}-H_{r}^{\dagger}}_\text{sp}dt. \label{eq: Hr purity bound}
\end{equation}
The advantage of the Liouville space is now apparent.
In contrast to the Hilbert space bound \eqref{eq: H_rho purity bound}, Eq.~\eqref{eq: Hr purity bound} was derived without using the triangle identity for the integrand and without the submultiplicativity property  of the norm. We demonstrate in the Appendix that, as a result, the Liouville space bound is always better than the Hilbert space bound for the case of a pure dephasing channel, $[A_{k},A_{k}^{\dagger}]=0$. In the more
general case, $[A_{k},A_{k}^{\dagger}]\neq0$, we have observed numerically that
this property holds true for randomly generated operators $A_k$, but were not able to show it analytically in full generality. Another argument in favor of the Liouville space bound can be found by looking at many-particle or many-level systems.  Let us consider $M$ independent particles subjected to the same  dynamics. Since the system is in a product state at all times, the log-purity scales like $ \ln {\mc P}_M \sim M \ln{\mc P}_1$. The Liouville space bound \eqref{eq: Hr purity bound} then exhibits the correct $M$ scaling, $ d{\mc P}_M/dt \leq M||H_r-H_r^\dagger||_\text{sp}$, in contrast to the Hilbert space bound  \eqref{eq: H_rho purity bound}, $d{\mc P}_M/dt \leq M 2^{M-1} ||A||^2_2$, which becomes worse with increasing  $M$.  In addition, for a single $N$-level dephasing channel with eigenvalues $\lambda_j(A)=\exp(i \varphi_j)$, we find that the purity speed in Liouville space is limited by  $\text{max}|\lambda_i-\lambda_j|^2 \leq 4 $ (see Appendix), while it increases with $N$, $\text{max}|\lambda_i-\lambda_j| \leq 4N$, in Hilbert space, thus badly overestimating the purity value for large $N$. 

The purity bound \eqref{eq: Hr purity bound} may be further significantly improved  in the following way. 
Let $\ket{r_{s}}$ be a specific solution of the quantum evolution $i\partial_t\ket{r_{s}}=H_{r}\ket{r_{s}}$. We define the deviation
vector as $\ket{r_{D}}=\ket r-\ket{r_{s}}$, and the corresponding  purity deviation as
$\mc P_{D} =\braket{r_{D}}{r_{D}}$.  The purity deviation has a simple geometrical meaning as the square of the Euclidean distance, $\text{tr}[(\rho-\rho_s)^2]$, between the  states $\rho$ and $\rho_s$ (the regular purity is the distance to the origin $\rho_s=0$). By taking the time
derivative of $\mc P_{D}$ and reiterating the above derivation, we readily find,
\begin{equation}
\left|\ln\frac{\mc P_{D}(t_f)}{\mc P_{D}(t_i)}\right|\le\int_{t_i}^{t_f}\nrm{H_{r}-H_{r}^{\dagger}}_\text{sp}dt ,\label{eq: P dev bound}
\end{equation}
where the purity $\mc P$ has  now been replaced by the  purity deviation $\mc P_{D}$.  While Eq.~\eqref{eq: P dev bound} is valid for all vectors $\ket{r_{s}}$, it becomes particularly useful when $\ket{r_{s}}$ is given by the steady state,  $i\partial_t\ket{r_{s}}=0$. The benefit of the  replacement $\mc P \rightarrow \mc P_{D} $ is that only the part
of the purity that changes in time is taken into account. The  purity deviation bound \eqref{eq: P dev bound} has the remarkable property that it may be tight at all times and all states for purely dephasing qubit channel (see below). We furthermore stress that the norm $\nrm{H_{r}-H_{r}^{\dagger}}_\text{sp}$ can be determined from the measurable Lindblad operators $A_k$.

\begin{figure}
\includegraphics[width=8.4cm]{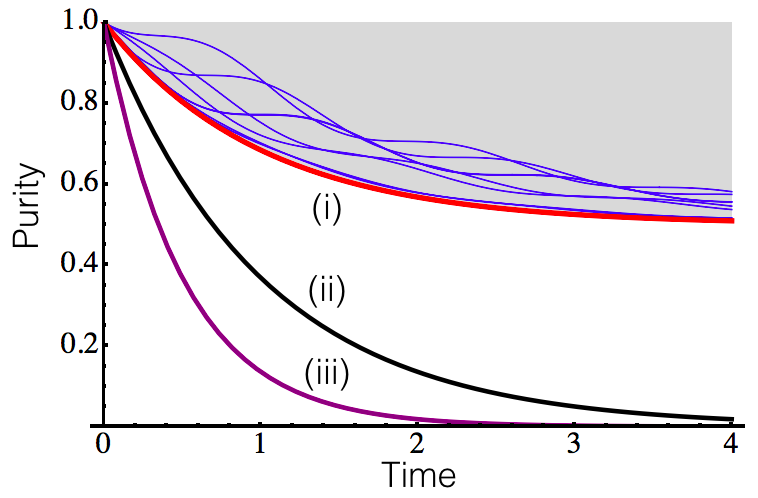}

\protect\caption{Purity bounds for  a qubit dephasing channel with $H_\rho=\sigma_x$ and $A= \sigma_z$ in a pure state at $t_i=0$. The   bound \eqref{12} (curve i) based on the purity deviation bound \eqref{eq: P dev bound} is tighter than the purity bound in Liouville space (\ref{eq: Hr purity bound}) (curve ii) and the corresponding bound in Hilbert space (\ref{eq: H_rho purity bound}) (curve iii). The bound (12) clearly delimits the region of allowed purities (blue lines obtained for various random initial conditions).}
\end{figure}
\textit{Applications.} We next apply our results to the case of a purely dephasing channel and to a problem of quantum coherent control in the presence of noise.

(a) Dephasing channel.  We first consider, for simplicity,  a single dephasing channel for a two-level system described by $H_{\rho}=0$ and
 one nonzero Lindblad operator that satisfies $[A,A^{\dagger}]=0$ \cite{nie00}. Without loss of generality, we  assume that the operator $A$ is traceless \cite{sau14}. In this situation, the Hilbert space bound (\ref{eq: H_rho purity bound})  takes a minimal value that is exactly two times larger compared to the tighter Liouville space bound (\ref{eq: Hr purity bound}). For  instance, for $A=\sigma_{z}$ and an initial density matrix of the  form $\rho({t_i})=\{\{a,b\},\{b^{*},1-a\}\}$, we find $|\ln {\mc P(t_f)}/{\mc P{(t_i)}}|\le2(t_f-t_i)$ in Hilbert space and $|\ln[{\mc P(t_f)}/{\mc P{(t_i)}}]|\le t_f-t_i$ in Liouville space. Remarkably,  the  purity deviation bound \eqref{eq: P dev bound}  is tight  at all times in this case: We choose $\rho_s$ to be the steady state given by the fully mixed state $\rho_{s}=\{\{a,0\},\{0,1-a\}\}$ \cite{rem1},  and  obtain the equality $|\ln[{\mc P_{D}(t_f)}/{\mc P_{D}(t_i)}]|=|\ln[{2b^{2}e^{-t_f}}/{(2b^{2}e^{-t_i})}]|=t_f-t_i$ which is exactly equal to the right hand side of \eqref{eq: P dev bound}.  We are not aware of the existence of any other tight speed limit for open quantum systems. In a large Hilbert space and in the presence of  multiple dephasing operators, the
purity deviation bound \eqref{eq: P dev bound} will be tight for initial conditions that populate 
the steady state and the fastest decaying mode exclusively.
   
Another merit of the purity deviation approach is that it can be used to get a better bound on the purity change for a general  dephasing channel in a $N$-level systems . Setting $\rho_s$  to be the fully mixed state (which always corresponds to a steady state in a dephasing dynamics) yields $\text{tr}[(\rho-\rho_s)^2]=\text{tr}[\rho^2]-1/N$. Using this in Eq. \eqref{eq: P dev bound} we obtain,
\begin{equation}
\mc P(t_f)\ge\frac{1}{N}+\left(\mc P(t_i)-\frac{1}{N}\right)\exp\left[{-\int_{t_i}^{t_f}\nrm{H_{r}-H_{r}^{\dagger}}_\text{sp}dt}\right]. \label{12}
\end{equation}
This equation, valid for any dephasing channel, is always better compared to Eq. \eqref{eq: Hr purity bound}.

Let us  illustrate the above results for a qubit dephasing channel with $H_\rho=\sigma_x$ and $A= \sigma_z$. Figure 1 shows the purity (blue lines) for various pure random initial conditions. We first observe that the  speed limit in Liouville space (\ref{eq: Hr purity bound}) (curve ii) is tighter than the bound obtained in Hilbert space (\ref{eq: H_rho purity bound}) (curve iii). We further note that the purity  bound \eqref{13} derived from the purity deviation bound \eqref{eq: P dev bound} is significantly better than the other two bounds. While it is not tight at all times for all initial states, it clearly  delimits the regime of allowed purity values.  

(b) Coherent control.
Coherent control,  in particular optimal control theory, is a powerful method for controlling dynamical processes in quantum mechanics \cite{sha03}. The technique deals with finding the time-dependent Hamiltonian necessary to implement a certain state transformation under  given restrictions
(e.g. the available operators in a Hamiltonian) \cite{pal02,kha11,sch05,bri10}. In Ref.~\cite{KallushNJPcohconNoise}
the effects of noise  in the control amplitude, coming from external degrees of freedom, were explored. In particular,
bounds on the minimal purity loss at the end of the evolution were
derived. These bounds are important to determine whether a quantum system that is fully controllable when isolated \cite{cla83,ram95}, remains fully controllable when coupled to an environment. We shall use our approach to put an upper
bound on the accumulated control dephasing noise. In a coherent control
setup the evolution of the density matrix $\rho$ is given by the Markovian master equation (\ref{eq: Lindblad eq.})
where $H_{\rho}=H_{0}+\sum_{k=1}^{M}f_{k}(t)H_{k}$ is the noise-free
control Hamiltonian and $A_{k}=n_{k}(t)^{1/2}H_{k}$ are dephasing terms
that arise from the noise in the control fields $f_{k}$ ($n_{k}\ge0$ without
loss of generality). Since the Lindblad operators $A_k$  are Hermitian in this case, Eq.~\eqref{eq: Lindblad eq.} describes a (multichannel) dephasing problem, according to the Lidar-Shabani-Alicki criterion \cite{LidarAlickiPurityDecrease}.

Employing Eq.~\eqref{12} for an initial pure state, $\mc P(t_i)=1$, we find,
\begin{equation}
\mc P(t_f)\ge\frac{1}{N}+\frac{N-1}{N}\exp\left[{-\int_{t_i}^{t_f}\nrm{H_{r}-H_{r}^{\dagger}}_\text{sp}dt}\right]. \label{13}
\end{equation}
Expression \eqref{13} gives a general lower bound on the final purity  in coherent control problems in terms of the spectral norm of the skew-Hermitian part $||{H_{r}-H_{r}^{\dagger}}||_\text{sp}$. It is always larger than $1/N$ (as it should) and larger than $\exp[{-\int_{t_i}^{t_f}||{H_{r}-H_{r}^{\dagger}}||_\text{sp}dt}$
predicted by the purity bound (\ref{eq: Hr purity bound}). 
It is sometimes possible to obtain even better results by not applying the triangle inequality. Let us consider the case of a single qubit with dephasing operators given by  the Pauli matrices $A_{k}=\sigma_{k}$. This example  also illustrates the concrete evaluation of the spectral norm of the skew-Hermitian part as a function  of the Lindblad operators.
When using the triangle  inequality in Liouville space, we obtain the inequality $-\ln\mc P(t_{f})\le \int_{t_i}^{t_f}\sum_{k}\left|n_{k}(t)\right|$.
However, the spectral norm can be here calculated analytically
without the triangle inequality to read $||{H_{r,k}-H_{r,k}^{\dagger}}||_\text{sp}=\sum_{k}\left|n_{k}(t)\right|-{\text{min}}_k\left|n_{k}(t)\right|$. Assuming that the noise amplitudes are all equal to $n_{0}$, we find a bound, $\mc P(t_{f}) \geq \exp(-3n_{0}(t_f-t_i))$, with the triangle inequality that is worse than the bound, $\mc P(t_{f}) \geq \exp(-2n_{0}(t_f-t_i))$, obtained  without. A similar calculation can be done for Eq. \eqref{12}.

\begin{figure}
\includegraphics[width=8.4cm]{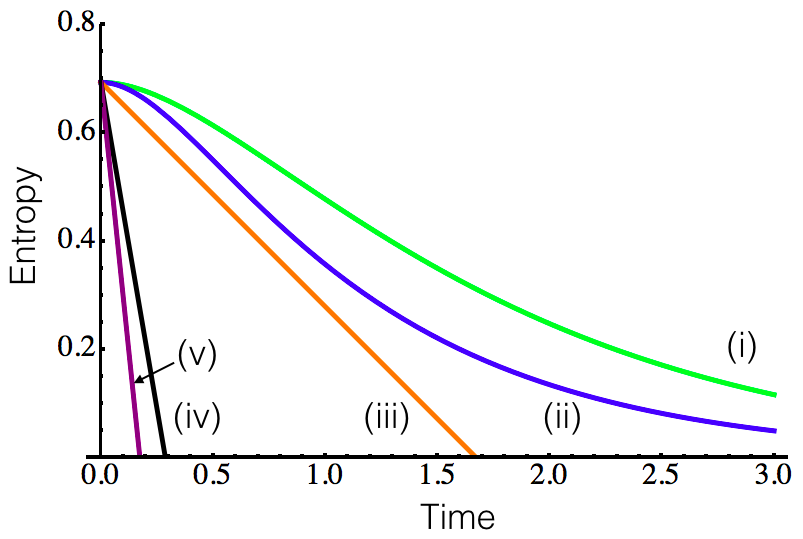}

\protect\caption{ Entropy bounds  for a qubit decay channel with $A=\sigma^{-}=(\sigma_{x}-i\sigma_{y})/2$ in a fully mixed state at $t_i=0$. The bound \eqref{14}  (curve iii) is better than the entropy bounds based on Eq.~(\ref{eq: Hr purity bound}) in Liouville space (curve iv) and  on Eq.~(\ref{eq: H_rho purity bound}) in Hilbert space (curve v). Curve i and ii respectively correspond to the exact  entropy $S(t)$ and exact log-purity $-\ln  \mc P(t)$.}
\end{figure}

\textit{Bound on entropy reduction.}
We have so far treated purity decrease (pure dephasing) and purity increase (cooling) on the same footing. However, we may obtain better bounds for cooling  processes by  replacing
 $\nrm{H_{r}-H_{r}^{\dagger}}_\text{sp}$ by 
$\text{max}(-i\, \text{eig}(H_{r}-H_{r}^{\dagger}))$ in expression (\ref{eq: Hr purity bound}) for the speed limit in Liouville space. This is no longer a norm and 
is valid only when purity is growing. Yet, when valid, it can be
significantly better than the spectral norm bound, as shown in Fig.
2 for the simple decay channel characterized by $A=\sigma^{-}=(\sigma_{x}-i\sigma_{y})/2$.  We may moreover obtain a bound on the decrease of the  von Neumann entropy with  the help of the  Jensen inequality, 
$S=-\text{tr}\rho\ln\rho\ge-\ln\mc P(t)$, and Eq.~\eqref{eq: Hr purity bound},
\begin{equation}
S(t_f)\ge-\ln\mc P(t_i)-\int_{t_i}^{t_f}\max[-i\: \text{eig}(H_{r}-H_{r}^{\dagger})]dt. \label{14}
\end{equation}
For  given initial and final entropies (or purities), expression \eqref{14} provides a bound on  how fast (minimal time interval $t_f-t_i$) a quantum system can be cooled in terms of the generators of the open dynamics. The appearance of the entropy instead of the purity establishes an important link to thermodynamics and highlights its relevance for the investigation of e.g. quantum heat engines.

\textit{Conclusions.}
We have derived state-independent quantum speed limits for the purity and the entropy of Markovian open quantum systems. We have obtained increasingly tighter bounds by considering the Liouville space instead of the usual Hilbert space. We have additionally shown that these bounds can be significantly improved by introducing a purity deviation obtained by subtraction of the steady-state contribution. We have finally emphasized the usefulness of these results  for the investigation of decoherence and thermalization processes in general, and applied them to concrete problems of  dephasing, noisy coherent control and cooling.

\section*{Appendix}
We show in the following that  the Liouville space bound (\ref{eq: Hr purity bound}) is tighter than the Hilbert space bound (\ref{eq: H_rho purity bound}) for the case of a dephasing channel. We begin by  considering  a single dephasing channel described, without loss of generality, by $H_{\rho}=0$ and
only one nonzero Lindblad operator that satisfies $[A,A^{\dagger}]=0$. The operator $A$ is unitary diagonalizable and has orthogonal eigenvectors
under the standard inner product. Hence, it uniquely defines 
an orthonormal basis through its eigenvectors. A density matrix
that is diagonal in this basis will  commute with $A$ and $A^{\dagger}$,
and lead to $L(\rho)=0$. Dephasing will therefore take
place in the eigenbasis of $A$. Without loss of generality
this dephasing channel can then be studied by taking $A$ to be a complex
diagonal matrix. Using the Hamilton superoperator definition (\ref{eq: Hr form}), we find that $H_{r}-H_{r}^\dagger$
is diagonal and therefore the eigenvalues are equal to the diagonal elements. By explicit calculation we find $\text{eig}($$H_{r}-H_{r}^{\dagger})=-i\{\left|\lambda_{i}-\lambda_{j}\right|^2\}_{i,j=1}^{N}$ where $\lambda_{i}$  are the eigenvalues of A. For diagonal matrices the spectral norm is the largest matrix element (in absolute value) and therefore:

\begin{equation}
\nrm{H_{r}-H_{r}^{\dagger}}_\text{sp} =\text{max}\left|\lambda_{i}-\lambda_{j}\right|^2 \le  4\nrm A_2^{2},\label{eq: A orthog ineq}
\end{equation}
where we used $\left|\lambda_{i}-\lambda_{j}\right|^2 \le (2\text{max}\left|\lambda_{i}\right|)^2\le  4\nrm A_2^{2}$. The speed limit in Liouville space  (\ref{eq: Hr purity bound}) is hence always better (or equal)
than the one in Hilbert space (\ref{eq: H_rho purity bound}) for
this type of dephasing evolution. 

The above proof can be easily extended to  a general dephasing channel with multiple
dephasing operators $[A_{k},A_{k}^{\dagger}]=0$.
We begin by applying the Hilbert space bound (\ref{eq: H_rho purity bound}) to an initially pure state to obtain $\left|\log\mc P(t_f)/\mc P(t_i)\right|\le4\int_{t_i}^{t_f}\sum_{k}\nrm{H_{k}}_{2}^{2}dt$.
On the other hand, using the Liouville space bound (\ref{eq: Hr purity bound}), we have, $\left|\log\mc P(t_f)/\mc P(t_i)\right| \le\int_{t_i}^{t_f}||\sum_{k}(H_{r,k}-H_{r,k}^{\dagger})||_\text{sp}dt$.
Using the triangle inequality $||\sum_{k}(H_{r,k}-H_{r,k}^{\dagger})||_{\text{sp}}\le\sum_{k}||{H_{r,k}-H_{r,k}||^{\dagger}}_{\text{sp}}$
and applying the inequality (\ref{eq: A orthog ineq}) for each $H_{r,k}$, we find
that the Liouville space bound is always tighter than the Hilbert space
bound for any multichannel dephasing $[A_{k},A_{k}^{\dagger}]=0$.

We finally mention that  in the
special case where the operator $A$ is Hermitian (e.g. in coherent control), the
log-purity speed bound (\ref{eq: Hr purity bound}) simplifies to,
\begin{equation}
\left|d_t(-\ln\mc P)\right|_{A=A^{\dagger}} \le  \Delta_{A}^{2}, \label{16}
\end{equation}
where $\Delta_{A}=\text{max}[\text{eig}(A)]-\text{min}[\text{eig}(A)]$  is the spectral gap of the operator.
Expression \eqref{13} is similar to the result, $|d_t\theta|\le (E_\text{max}-E_\text{min})/2$, obtained in the unitary case \cite{uzdin100evoSpeed}  ($E_\text{max/min}$ are the maximal/minimal instantaneous
eigenvalues of the Hamiltonian) \cite{com}. 


\begin{thebibliography}{99}
\bibitem{bre07} H.-P. Breuer and F. Petruccione, \textit{The Theory of Open Quantum Systems}, (Oxford, Oxford, 2007).
\bibitem{nie00} M. A. Nielsen and I. L. Chuang, \textit{Quantum Computation and Quantum Information}, (Cambridge, Cambridge, 2000).
\bibitem{sha03} M. Shapiro and P. Brumer, \textit{Principles of the Quantum Control of Molecular Processes}, (Wiley, New York, 2003).
\bibitem{wan13} X. Wang, S. Vinjanampathy, F. W. Strauch, and K. Jacobs, Phys. Rev. Lett. \textbf{110}, 157207 (2013).
\bibitem{kos14} R. Kosloff and A. Levy, Annu. Rev. Phys. Chem. \textbf{65}, 365 (2014).
\bibitem{fle73} G. N. Fleming, Nuovo Cimento A {\bf 16}, 232 (1973).
\bibitem{bha83} K. Bhattacharyya, J. Phys. A {\bf 16}, 2993 (1983).
\bibitem{ana90} J. Anandan and Y. Aharonov, Phys. Rev. Lett. \textbf{65}, 1697 (1990).
\bibitem{vai91} L. Vaidman, Am. J. Phys. {\bf 60}, 182 (1991).
\bibitem{uff93} J. Uffink, Am. J. Phys. {\bf 61}, 935 (1993).
\bibitem{bro03} D. C. Brody, J. Phys. A \textbf{36}, 5587 (2003).
\bibitem{man45} L. Mandelstam and I. Tamm, J. Phys. (USSR) \textbf{9}, 249 (1945).
\bibitem{mar98} N. Margolus and L. B. Levitin, Physica D \textbf{120}, 188 (1998).
\bibitem{gio03} V. Giovannetti, S. Lloyd and L. Maccone, Phys. Rev. A \textbf{67}, 052109 (2003).
\bibitem{pfe93} P. Pfeifer, Phys. Rev. Lett. \textbf{70}, 3365 (1993).
\bibitem{pfe95} P. Pfeifer and J. Fr\"ohlich, Rev. Mod. Phys. \textbf{67}, 759 (1995).
\bibitem{def13} S. Deffner and E. Lutz,  J. Phys. A \textbf{46}, 335302 (2013).
\bibitem{tad13} M. M. Taddei, B. M. Escher, L. Davidovich, and R. L. de Matos Filho, Phys. Rev. Lett. \textbf{110}, 050402 (2013).
\bibitem{cam13} A. del Campo, I. L. Egusquiza, M. B. Plenio, and S. F.
Huelga, Phys. Rev. Lett. \textbf{110}, 050403 (2013).
\bibitem{def13a} S. Deffner and E. Lutz, Phys. Rev. Lett. \textbf{111}, 010402 (2013).
\bibitem{rod11} C. A. Rodriguez-Rosario, G. Kimura, H. Imai, and A. Aspuru-Guzik, Phys. Rev. Lett. \textbf{106}, 050403 (2011).
\bibitem{hut12} A. Hutter and S. Wehner, Phys. Rev. Lett. \textbf{108}, 070501 (2012).
\bibitem{LidarNormActionAndDistanceBound} D. A. Lidar, P. Zanardi, and K. Khodjasteh, Phys. Rev. A \textbf{78}, 012308 (2008).
\bibitem{uzdin100evoSpeed} R. Uzdin, U. G\"unther, S. Rahav, and N. Moiseyev, J. Phys. A \textbf{45}, 415304 (2012).
\bibitem{uzdinNHresources} R. Uzdin, J. of Phys. A \textbf{46}, 145302 (2013).
\bibitem{LidarAlickiPurityDecrease} D. Lidar, A. Shabani, and R. Alicki, Chem. Phys. \textbf{322}, 82 (2006).

\bibitem{bha97} R. Bhatia, \textit{Matrix Analysis}, (Springer, Berlin, 1997).
\bibitem{RenyiEntropyQM} M. M\"uller-Lennert, F. Dupuis, O. Szehr, S. Fehr, and M. Tomamichel, J. Math. Phys. \textbf{54},
122203 (2013).
\bibitem{chu97} I.L. Chuang and M.A. Nielsen, J. Mod. Opt. \textbf{44}, 2455 (1997).
\bibitem{chi01} A. M. Childs, I. L. Chuang, and D. W. Leung, Phys. Rev. A \textbf{64}, 012314 (2001).
\bibitem{how06} M. Howard, J. Twamley, C. Wittmann, T. Gaebel, F. Jelezko, and J. Wrachtrup, New J. Phys. \textbf{8}, 33 (2006).
\bibitem{myr05} S. H. Myrskog, J. K. Fox, M. W. Mitchell, and A. M. Steinberg, Phys. Rev. A \textbf{72}, 013615 (2005).
\bibitem{obr04} J. L. O'Brien, G. J. Pryde, A. Gilchrist, D. F. V. James, N. K. Langford, T. C. Ralph, and A. G. White, Phys. Rev. Lett. \textbf{93}, 080502 (2004).
\bibitem{nam05} Y. Nambu and K. Nakamura, Phys. Rev. Lett. \textbf{94}, 010404 (2005).
\bibitem{rie06} M. Riebe, K. Kim, P. Schindler, T. Monz, P. O. Schmidt, T. K. Korber, W. Hansel, H. Haffner, C. F. Roos, and R. Blatt, Phys. Rev. Lett. \textbf{97}, 220407 (2006).
\bibitem{nee08} M. Neeley, M. Ansmann, R. C. Bialczak, M. Hofheinz, N. Katz, E. Lucero, A. O'Connell, H. Wang, A. N. Cleland, and J. M. Martinis, Nat. Phys. \textbf{4}, 523 (2008).
\bibitem{bru08} M. Brune, J. Bernu, C. Guerlin, S. Del\'eglise, C. Sayrin, S. Gleyzes, S. Kuhr, I. Dotsenko, J. M. Raimond, and S. Haroche, Phys. Rev. Lett. \textbf{101}, 240402 (2008).
\bibitem{wan08} H. Wang, M. Hofheinz, M. Ansmann, R. C. Bialczak, E. Lucero, M. Neeley, A. D. O'Connell, D. Sank, J. Wenner, A. N. Cleland, and J. M. Martinis, Phys. Rev. Lett. \textbf{101}, 240401 (2008).
\bibitem{bov05} F. A. Bovino, G. Castagnoli, A. Ekert, P. Horodecki, C. M. Alves, and A. V. Sergienko, Phys. Rev. Lett. \textbf{95}, 240407 (2005).
\bibitem{du06} J. Du, P. Zou, X. Peng, D. K. L. Oi, L. C. Kwek, C. H. Oh, and A. Ekert, Phys. Rev. A \textbf{74}, 042319 (2006).
\bibitem{ada07} R. B. A. Adamson, L. K. Shalm, and A. M. Steinberg, Phys. Rev. A \textbf{75}, 012104 (2007).
\bibitem{eke02} A. K. Ekert, C. M. Alves, D. K. L. Oi, M. Horodecki, P. Horodecki, and L. C. Kwek, Phys. Rev. Lett. \textbf{88}, 217901 (2002).
\bibitem{muk95} S. Mukamel, \textit{Principles of Nonlinear Spectroscopy}, (Oxford, Oxford, 1995).
\bibitem{sau14} S. Sauer, C. Gneiting, and A. Buchleitner, Phys. Rev. A \textbf{89}, 022327 (2014).
\bibitem{rem1} There may be several steady states depending on the initial
state and one may choose one of them. For the considered dephasing channel, the second steady state is $\rho_{s}=\{\{1/2,0\},\{0,1/2\}\}$ (the purity bound is improved when using this steady state, but not tight). 
\bibitem{pal02} J. P. Palao and R. Kosloff, Phys. Rev. Lett. \textbf{89}, 188301 (2002).
\bibitem{kha11} M. Khasin and R. Kosloff, Phys. Rev. Lett. \textbf{106} 123002 (2011).
\bibitem{sch05} T. Schulte-Herbr\'uggen, A. Sp\"orl, N. Khaneja, and S.J. Glaser, Phys. Rev. A \textbf{72}, 042331 (2005).
\bibitem{bri10} C. Brif, R. Chakrabarti, and H. Rabitz, New J. Phys. \textbf{12}, 075008  (2010). 
\bibitem{KallushNJPcohconNoise} S. Kallush, M. Khasin, and R. Kosloff, New J. Phys. \textbf{16}, 015008 (2014).
\bibitem{cla83} J. Clark and T. Tarn, J. Math. Phys. \textbf{24}, 2608 (1983).
\bibitem{ram95} V. Ramakrishna, M.V. Salapaka, M. Dahlem, H. Rabitz, and A. Pierce, Phys. Rev. A \textbf{51}, 960 (1995).
\bibitem{com}  The spectral gap in Eq.~(13) is squared,  since $A$ appears in a quadratic form in the Lindblad master equation (2). 


\end{thebibliography}
\end{document}